\documentclass[prd, onecolumn, notitlepage, nofootinbib, floatfix]{revtex4-1}
\usepackage{amsmath}
\usepackage{graphicx}
\usepackage{dcolumn}
\usepackage{bm}
\usepackage{epsfig}
\usepackage{amssymb,latexsym,mathrsfs}
\usepackage{graphicx}
\usepackage{color}
\usepackage{centernot} 
\usepackage{cancel} 
\usepackage{hyperref}

\usepackage{float}

\hypersetup{
    colorlinks=true,
    linkcolor=red,
    citecolor=blue,
} 
\newcommand{\be}{\begin{equation}}
\newcommand{\ee}{\end{equation}}
\newcommand{\bs}{\begin{split}} 
\newcommand{\bea}{\begin{eqnarray}}
\newcommand{\eea}{\end{eqnarray}}
\newcommand{\mpl}{M^2_{\rm pl}} 
 
\newcommand{\gl}{G_{\rm light}} 
\newcommand{\gm}{G_{\rm matter}} 
\newcommand{\geff}{G_{\rm eff}} 
\newcommand{\alm}{\alpha_M} 
\newcommand{\alb}{\alpha_B} 
\newcommand{\alk}{\alpha_K} 
\newcommand{\al}{\alpha} 
\newcommand{\rhol}{\rho_\Lambda} 
\newcommand{\rhomo}{\rho_{m,0}} 
\newcommand{\olm}{\Omega_\Lambda} 
\newcommand{\oli}{\Omega_{\Lambda,i}}

\begin{document}

\title{Horndessence: $\Lambda$CDM Cosmology from Modified Gravity} 

\author{Eric V.\ Linder$^{1,2}$} 
\affiliation{
${}^1$Berkeley Center for Cosmological Physics \& Berkeley Lab, 
University of California, Berkeley, CA 94720, USA\\ 
${}^2$Energetic Cosmos Laboratory, Nazarbayev University, 
Nur-Sultan 010000, Kazakhstan
}

\begin{abstract} 
Rather than obtaining cosmic acceleration with a scalar field potential 
(quintessence) or noncanonical kinetic term (k-essence), we can do 
it purely through a modified gravity braiding of the scalar and metric, 
i.e.\ the $G_3$ Horndeski action term. 
Such ``Horndessence'' allows an exact $\Lambda$CDM cosmological 
expansion without any cosmological constant, and by requiring 
shift symmetry we can derive the exact form of $G_3$. We find that 
this route of deriving $G_3(X)$ leads to a functional form far 
from the usual simple assumptions such as a power law. 
Horndessence without any kinetic term or potential has the same 
number of parameters as $\Lambda$CDM and makes an exact 
prediction for the expansion history ($\Lambda$CDM) and 
modified gravity cosmic growth history; we show the viable 
gravitational strength $\geff(a)$ and growth rate $f\sigma_8(a)$. 
The simplest versions of 
the theory fail soundness criteria, but we learn  interesting lessons 
along the way, in particular about robust parametrization, and 
indicate possible sound extensions. 
\end{abstract}

\date{\today} 

\maketitle

\section{Introduction}

Current cosmic acceleration could be due to a cosmological constant, 
a constant vacuum energy density. Quintessence poses an alternative where 
there is a dynamical scalar field, rolling in a potential. Cosmic 
acceleration is also possible without a potential, through changing the 
kinetic structure of the scalar field, as in purely kinetic k-essence 
\cite{kess1,kess2,0705.0400,gubilin}. Here we explore using neither the potential 
nor kinetic structure, but modified gravity to deliver an effective 
cosmological constant, so the cosmic expansion is not only accelerating but 
identical to the $\Lambda$CDM cosmology. 

Alternatives are interesting and useful since 
numerous issues exist with employing a potential, or a cosmological constant, 
as they should be altered by quantum corrections. An effective cosmological 
constant, without any actual vacuum energy, is therefore an interesting idea 
to pursue, especially if we do 
so within the framework of a shift symmetric theory, which ameliorates many 
quantum corrections. 

Here we investigate the cosmology where Horndeski gravity acts like an 
effective cosmological constant, exploring its observational impact on 
the modified gravitational strengths for cosmic structure growth and light 
propagation and prediction for growth rate  measurable by redshift 
space distortions in galaxy surveys, 
the consequences for the form of the Horndeski functions, and theoretical 
soundness. Section~\ref{sec:eqs} lays out the basic equations of motion 
and classes of theories, while Section~\ref{sec:sols} derives the 
solutions and observational impacts. Section~\ref{sec:sound} treats 
the soundness of the theories and we conclude in Section~\ref{sec:concl}.

\section{How to Get a Cosmological Constant} \label{sec:eqs} 

The Horndeski action is 
\be 
S = \int d^{4}x \sqrt{-g} \,\Bigl[ G_4(\phi)\, R + K(\phi,X) -G_3(\phi,X)\Box\phi  + {\cal L}_m[g_{\mu\nu}]\,\Bigr]\,, 
\label{eq:action} 
\ee 
where $\phi$ is the scalar field, $X=-(1/2)g^{\mu\nu}\phi_\mu\phi_\nu$, 
$R$ is the Ricci scalar, ${\cal L}_m$ the matter Lagrangian, and 
$K$, $G_3$, $G_4$ the Horndeski terms. We will work within a 
Friedmann-Lema\^{i}tre-Robertson-Walker (FLRW) cosmology. Note that taking 
$G_5=0$ and $G_4=G_4(\phi)$ ensures the speed of gravitational wave 
propagation is equal to the speed of light. 

The equations of motion, writing $2G_4=\mpl+A(\phi)$, are 
\bea 
3H^2(\mpl+A)&=&\rho_m+2XK_X-K+6H\dot\phi XG_{3X} 
-2XG_{3\phi}-3H\dot\phi  A_\phi\label{eq:fried}\\ 
-2\dot H\,(\mpl+A)&=&
\ddot\phi\,(A_\phi-2XG_{3X})-H\dot\phi\,(A_\phi-6XG_{3X})
+2X\,(A_{\phi\phi}-2G_{3\phi})\notag\\ 
&\qquad&+2XK_X+\rho_m+P_m \label{eq:fulldh}\\ 
0&=&\ddot\phi\,\left[K_X+2XK_{XX}-2G_{3\phi}-2XG_{3\phi X}+6H\dot\phi(G_{3X}+XG_{3XX})\right]\notag\\ 
&\qquad&+3H\dot\phi\,(K_X-2G_{3\phi}+2XG_{3\phi X})-K_\phi\notag\\ 
&\qquad&+2X\,\left[K_{\phi X}-G_{3\phi\phi}+3G_{3X}(\dot H+3H^2)\right] 
-3A_\phi(\dot H+2H^2)\,.  \label{eq:fullddphi} 
\eea 

Thus the Horndeski and matter terms determine the cosmic expansion rate 
$H$ and scalar field evolution $\phi$. Note the mix of dependent variables: 
we have for example $G_3(\phi,X)$ but $\rho_m(a)$, $H(a)$. 
We must solve the coupled equations to figure out how they are related, i.e.\ 
$a(\phi,X(\phi))$. This can be quite involved except for the simplest
forms of the Horndeski functions, e.g.\ power laws. 

Here we will take a different approach, and specify not the Horndeski 
functions but the Hubble parameter $H$, in particular as exactly that 
of $\Lambda$CDM:  
\be 
3\mpl H^2=\rho_m+\rho_\Lambda=(3\mpl H_0^2-\rhol)\,a^{-3}+\rhol\,, 
\label{eq:hlam}
\ee 
where $H_0=H(a=1)$ and $\rhol$ is a constant (effective) energy density. 
From this we will attempt to derive the Horndeski 
functional forms and scalar field evolution. 

This approach cannot be done in general. Consider the effective dark  
energy density and pressure, 
\bea 
\rho_{\rm de}&=&2XK_X-K+6H\dot\phi XG_{3X}-2XG_{3\phi}-3H\dot\phi A_\phi 
\label{eq:rhode}\\ 
\rho_{\rm de}+P_{\rm  de}&=&
\ddot\phi\,(A_\phi-2XG_{3X})-H\dot\phi\,(A_\phi-6XG_{3X})
+2X\,(A_{\phi\phi}-2G_{3\phi})+2XK_X\,. \label{eq:rhopp} 
\eea 
For our effective cosmological constant, $\rho_{\rm de}+P_{\rm  de}=0$, 
but we have a mix of $\dot\phi(t)$ and the Horndeski functions which 
cannot be separately determined without further assumptions. 

Following the idea of quintessence, which uses only the kinetic function 
in the action Eq.~\eqref{eq:action}, 
(in the form $K=\dot\phi^2/2-V(\phi)$), or purely kinetic k-essence, which 
uses only the kinetic function $K(X)$, let us explore one Horndeski function 
at a time. If we try to use only $G_4(\phi)$, then the scalar field equation 
\eqref{eq:fullddphi} is merely $A_\phi(\dot H+2H^2)=0$; this is only valid 
for FLRW cosmology if $A_\phi=0$, i.e.\ $G_4$ is just the usual constant 
Planck mass squared (over 2). If we take $G_3(\phi)$, i.e.\ no dependence 
on $X$, this can be converted to k-essence \cite{2009.01720} (though not 
purely kinetic k-essence in general, rather $K(\phi,X)$). For the 
(noncanonical) kinetic term alone, the $\dot H$ equation~\eqref{eq:fulldh} 
gives simply $XK_X=0$ for an effective cosmological constant; then 
Eq.~\eqref{eq:fullddphi} implies $K_\phi=0$ too, so there is no such 
cosmological constant behavior solution. 

Thus we are left with $G_3(X)$. This also has the nice property of being 
shift symmetric, though we will relax that at times. 
Such an action -- though with a kinetic term -- has been used in 
kinetic gravity braiding dark energy \cite{kgb}, inflation \cite{ginfl}, 
and to address the original cosmological constant problem \cite{temper}. 

Now Eq.~\eqref{eq:rhopp} 
becomes 
\be 
0=-2XG_{3X}\,(\ddot\phi-3H\dot\phi)\,,  \label{eq:gphi} 
\ee 
with solution 
\bea 
\ddot\phi&=&3H\dot\phi\\ 
\dot\phi&=&\dot\phi_i\,\left(\frac{a}{a_i}\right)^3\,. \label{eq:dphia} 
\eea 
With the solution for $\dot\phi(a)$ (and implicitly $\phi(a))$, we can 
then use Eq.~\eqref{eq:rhode} (or equivalently Eq.~\ref{eq:fullddphi}) 
to determine $G_3(X)$. This demonstrates the inverse path of the 
usual construction -- starting with $\Lambda$CDM and deriving $G_3(X)$, 
rather than  
starting with the Horndeski functions and deriving the cosmic expansion 
evolution $H(a)$. 

Interestingly, the scalar field motion is independent of the form of 
$G_3$; the kinetic energy simply grows with scale factor as $X\sim a^6$ 
(recall no potential was introduced). This is like a time reversed 
version of ``skating'' \cite{skate1,skate2}, where a field on a flat 
potential glides with diminishing kinetic energy $X\sim a^{-6}$;  
here the modified gravity speeds up the field rather than the field slowing 
due to Hubble friction ultimately to a stop (cosmological constant).

\section{Deriving $G_3$ and Cosmological Influences} \label{sec:sols} 

Let us proceed to derive $G_3(X)$ without allowing any other Horndeski term, 
including the kinetic term $K(X)$. While we hold to no  $K(X)$, we will 
also explore the effect of having a potential as part of $K$ (recall that 
for quintessence, for example, $K=X-V$). 

\subsection{Zero Potential} \label{sec:nov} 

In the absence of any potential, Eq.~\eqref{eq:rhode} becomes 
\be 
6\sqrt{2}\,HX^{3/2}G_{3X}=\rhol\,. \label{eq:geqnov} 
\ee 
Using Eq.~\eqref{eq:hlam} and recalling that $\rhol$ is a constant, the 
solution is 
\be 
G_3(X)=G_3(X_i)-\frac{\rhol}{\rhomo}\sqrt{\frac{2\mpl}{3X_i a_i^{-6}}}\, 
\left[\sqrt{\rhomo(X/X_i)^{-1/2}+\rhol}-\sqrt{\rhomo+\rhol}\right]\,. 
\ee 
Recall that $\rhomo=3\mpl H_0^2-\rhol$. It is interesting that even 
in this extremely simple situation of a single Horndeski function of 
a single variable, the functional form corresponding to the concordance 
cosmology is not trivial. That is, if one attempted to parametrize the 
Horndeski functions a priori, one might not guess forms like 
$\sqrt{bX^{-1/2}+c}=\sqrt{b}\,X^{-1/4}\sqrt{1+(c/b)X^{1/2}}$. 
Interestingly, a similar complexity to match a $\Lambda$CDM background 
was found for the reconstruction 
of $K(X)$ for a simple physical system with kinetic gravity braiding 
\cite{kgb2011}, for $f(R)$ gravity \cite{multamaki,odintsov}, and 
for some DHOST theories \cite{motohashi}. Our result agrees 
with \cite{nesseris}, who also analyzed the no potential case.

\subsection{Linear Potential} \label{sec:linv} 

We  can make a slight elaboration by adding a linear potential. Such a potential 
$V=\lambda^3\phi$ is shift symmetric and has been 
used for cosmic acceleration in inflation and dark energy \cite{linpoti,linpotde}. 
We keep $K_X=0$ but now $K_\phi=-\lambda^3$. 

This does not affect Eq.~\eqref{eq:gphi} or its solution 
Eq.~\eqref{eq:dphia} for $\dot\phi(a)$, but does impact the solution for 
the functional form $G_3(X)$. Equation~\eqref{eq:geqnov} now 
becomes 
\be 
6\sqrt{2}\,HX^{3/2}G_{3X}=\rhol-\lambda^3\phi\,. \label{eq:geqlinv} 
\ee 

To solve this we need $\phi(X(a))$. The integral of Eq.~\eqref{eq:dphia} 
yields 
\bea 
\phi(a)&=&\phi(a_i)+\dot\phi_i a_i^{-3}\int dt\,a^3\\ 
&=&\phi(a_i)+\dot\phi_i a_i^{-3}\int_{a_i}^a dA \,A^2 
\frac{3\mpl}{\sqrt{\rhomo A^{-3}+\rhol}}\\ 
&=&\phi(a_i)-\dot\phi_i a_i^{-3} \frac {\rhomo}{2\rhol} \sqrt{\frac{\mpl}{3\rhol}} 
\left(\ln\left[\frac{\left(\sqrt{3\mpl H^2/\rhol+1}\right)\left(\sqrt{3\mpl H_i^2/\rhol-1}\right)}{\left(\sqrt{3\mpl H^2/\rhol-1}\right)\left(\sqrt{3\mpl H_i^2/\rhol+1}\right)}\right]-\frac{2\sqrt{3\mpl H^2/\rhol}}{3\mpl H^2/\rhol-1}+\frac{2\sqrt{3\mpl H_i^2/\rhol}}{3\mpl H_i^2/\rhol-1}\right) \label{eq:phih}\\ 
&=&\phi(a_i)-\dot\phi_i a_i^{-3} \frac{\rhomo}{2\rhol} \sqrt{\frac{\mpl}{3\rhol}} 
\left(\ln\left[\frac{\left(1+\sqrt{\olm}\right)\left(1-\sqrt{\oli}\right)}{\left(1-\sqrt{\olm}\right)\left(1+\sqrt{\oli}\right)}\right]-\frac{2\sqrt{\olm}}{1-\olm}+\frac{2\sqrt{\oli}}{1-\oli}\right)\,. \label{eq:phio} 
\eea 
Recall that in terms of $X$, 
\bea 
H^2(X)&=&\frac{\rhomo\,(X/X_i)^{-1/2}+\rhol}{3\mpl} \label{eq:hasx}\\ 
\olm(X)&=&\left[1+\frac{\rhomo}{\rhol} a_i^{-3}\left(\frac{X}{X_i}\right)^{-1/2}\right]^{-1}\,. \label{eq:olmasx} 
\eea 

Using that 
\be  
G_{3X}=\frac{dG_3}{dH^2}\,\frac{dH^2}{dX}=\frac{dG_3}{dH^2}\,X^{-3/2}\frac{-\rhomo\sqrt{X_i a_i^{-3}}}{\mpl}\,, 
\ee 
gives to $G_3(a)$  a contribution from the second term of 
Eq.~\eqref{eq:geqlinv} as an integral of the form 
\be 
\int dH\,\phi=\sqrt{\frac{\rho}{3\mpl}}\,\left[(y+1)\ln(y+1)-(y-1)\ln(y-1)-(y_i+1)\ln(y_i+1)-(y_i-1)\ln(y_i-1)-\ln\frac{y^2-1}{y_i^2-1}\right]\,, 
\ee  
where $y=\sqrt{1/\olm}=\sqrt{3\mpl H^2/\rhol}$. 
The form of $G_3$ is then 
\bea 
G_3(X)&=&G_3(X_i)-\frac{\rhol}{\rhomo}\sqrt{\frac{2\mpl}{3X_i a_i^{-6}}}\, 
\left[\sqrt{\rhomo(X/X_i)^{-1/2}+\rhol}-\sqrt{\rhomo+\rhol}\right]\notag\\ 
&+&\frac{\lambda^3\mpl\sqrt{2}}{\rhomo\sqrt{X_ia_i^{-6}}} 
\left\{\phi_i\,(H-H_i)+\dot\phi_i a_i^{-3}\frac{\rhomo}{2\rhol}\sqrt{\frac{\mpl
}{3\rhol}}\left[\ln\frac{\sqrt{1/\oli}+1}{\sqrt{1/\oli}-1}-\frac{2\sqrt{\oli}}{1/\oli-1}\right]\,(H-H_i)\right.\notag \\ 
&\qquad&-\dot\phi_i a_i^{-3}\,\frac{\rhomo}{6\rhol}\,\left[(\sqrt{1/\olm}+1)\ln(\sqrt{1/\olm}+1)-(\sqrt{1/\oli}+1)\ln(\sqrt{1/\oli}+1)\right.\notag\\ 
&\qquad&\quad-\left.\left.(\sqrt{1/\olm}-1)\ln(\sqrt{1/\olm}-1)+(\sqrt{1/\oli}-1)\ln(\sqrt{1/\oli}-1)+\ln\frac{1/\oli-1}{1/\olm-1}\right]\right\}\,. \label{eq:g3full}  
\eea 

For the full glory (see Appendix~\ref{sec:apxfull}), one would expand the notation using Eqs.~\eqref{eq:hasx} and \eqref{eq:olmasx} to show 
$G_3(X)$ for this very simple model, with only a linear potential and no $K(X)$ or $G_4(\phi)$, 
or $G_3(\phi)$ -- we emphasize that $G_3(X)$ in the above expression,  with $\phi(X)$ from 
Eq.~\eqref{eq:phio} and $\dot\phi=(2X)^{1/2}$ in Eq.~\eqref{eq:dphia}, gives the $\Lambda$CDM 
solution to the Friedmann equations. 
It seems quite unlikely that someone starting from $G_3(X)$ would have chosen such a functional form; 
we see that assuming simple, e.g.~power law, forms should have no expectation of capturing 
accurately the detailed physics of a cosmic expansion near $\Lambda$CDM.

\subsection{Cosmological Impact} 

Apart from the expansion history, modified gravity affects the growth of  
cosmic structure through the time varying effective gravitational strength -- 
the gravity history. Generally there are two gravitational strengths, seen 
in two modified Poisson equations: those relating the time-time metric 
potential to the matter density perturbation and relating the sum of the 
time-time and space-space metric potentials to the matter density 
perturbation. These can be abbreviated $\gm$ and $\gl$ respectively. 

They are related to the Horndeski functions and their derivatives, but 
a compact notation in terms of property functions $\al_i$ was given by 
\cite{belsaw}. For computing the gravitational strengths we need 
\bea 
\alm=\frac{d\ln 2G_4}{d\ln a}\\ 
\alb=\frac{\dot\phi XG_{3X}}{HG_4}\,, 
\eea 
for our class of modified gravity, where $\alpha_T=0$ since we took 
the speed of gravitational waves to be the speed of light. As our 
$2G_4=\mpl$ is constant, we also have $\alm=0$. Thus our theory is 
part of the No Run Gravity class \cite{nrg}, and $\gm=\gl$ (thus 
there is no gravitational slip) so we will 
simply refer to $\geff$. 

Evaluating our solutions for $G_3$, we have 
\be 
\alb(a)=\olm(a)\,\left(1-\frac{\lambda^3 \phi}{\rhol}\right)\,, \label{eq:alb} 
\ee 
where $\phi(a)$ is given by Eq.~\eqref{eq:phih} or \eqref{eq:phio}. 
Note the beautifully simple form for the no potential case: 
$\alb(a)=\olm(a)$. This would go to zero in the early universe, 
indicating an unmodified general relativity, and freeze to unity in 
the de Sitter future. 

The gravitational strength for the No Run Gravity class is 
\be 
\geff=\frac{\alb+\alb'}{\alb-\alb^2/2+\alb'}\ , 
\ee 
where $\geff$ is in units of Newton's constant and 
a prime denotes $d/d\ln a$. For the no potential and 
linear potential models this yields 
\be 
\geff=\frac{\olm(4-3\olm)(1-\lambda^3\phi/\rhol)-\lambda^3\dot\phi_i a_i^{-3}(\olm/\rhol)(a^3/H)}{\olm(4-3\olm)(1-\lambda^3\phi/\rhol)-\lambda^3\dot\phi_i a_i^{-3}(\olm/\rhol)(a^3/H)-(1/2)\olm^2(1-\lambda^3\phi/\rhol)^2}\ . 
\ee 
At early times the $\olm^2$ term from $\alb^2$ is negligible 
compared to the linear $\alb$ term, and $\geff\to1$. Again this 
indicates general relativity holds in the early universe. At late 
times, $\geff\to2$ in the no potential case ($\lambda^3=0$) 
and $\geff\to0$ in the linear potential case (not surprising, as 
the field rolls to infinity and $G_3$, and hence $\alb$, must grow to match it in 
order to keep $\rho_{\rm de}=\rhol$ for the $\Lambda$CDM background).  
For the no potential case the gravitational strength takes a 
simple form, 
\be 
\geff=\frac{1-(3/4)\olm}{1-(7/8)\olm}\ . \qquad{\rm [No\ potential]} 
\ee

\begin{figure}[htb!]
\centering 
\includegraphics[width=0.6\columnwidth]{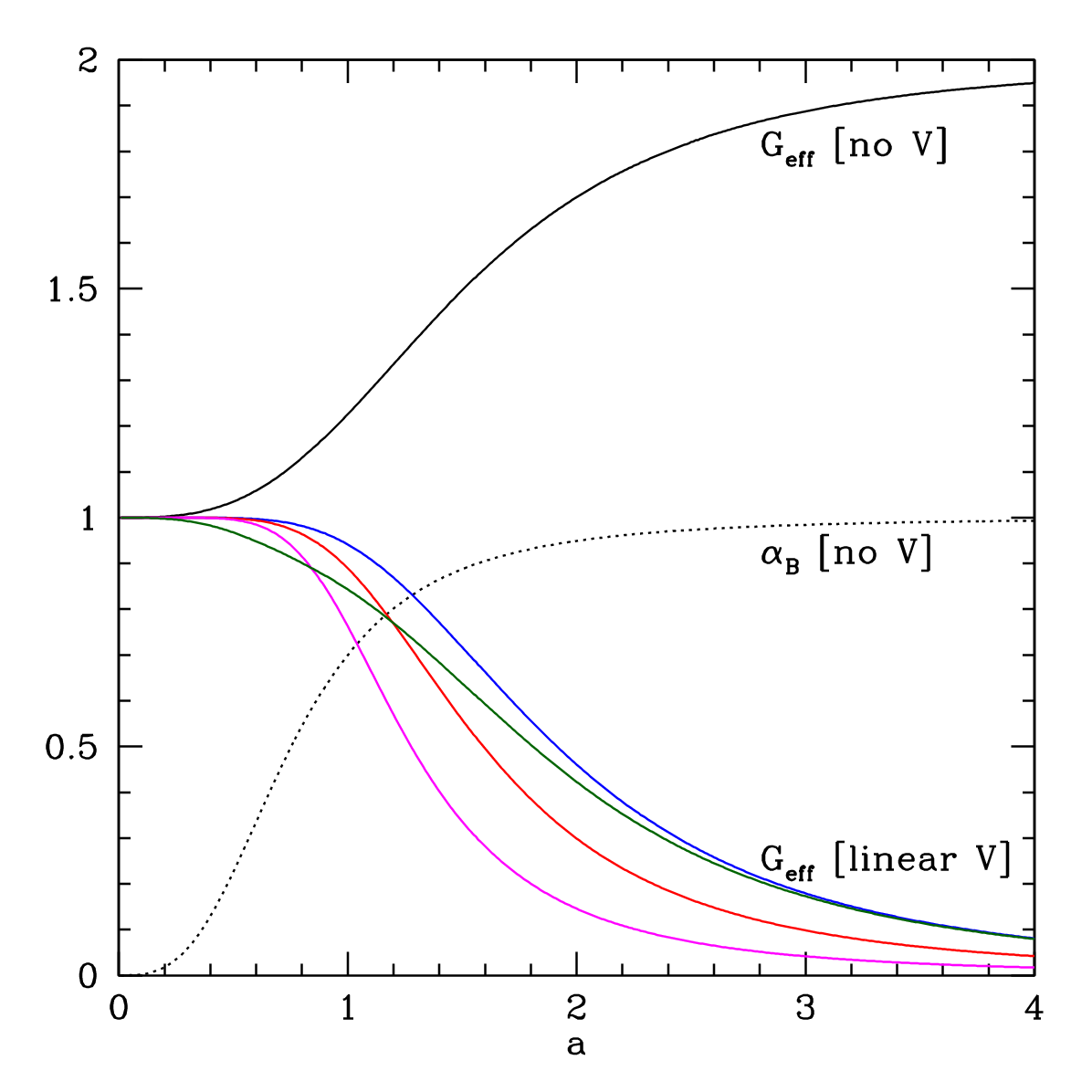}
\caption{The effective gravitational strength $\geff$ is plotted 
for the no potential and various linear potential models (solid 
curves). The dotted curve is $\alb$ for the no potential case; 
for the linear potential cases $\alb$ goes very negative near 
the present. The linear potential cases have 
$(\kappa_i,\kappa_0-\kappa_i)=(1,1)$, $(1, 2)$, $(1, 5$), $(2, 1$) for the 
blue, red, magenta, dark green curves respectively, from 
top to bottom at $a=0.7$. 
} 
\label{fig:geff} 
\end{figure}

Figure~\ref{fig:geff} plots the braiding property function $\alb$ 
and the gravitational strength $\geff$ for the no potential and 
linear potential cases. For the linear potential model there are 
two parameters, essentially the initial energy density $\lambda^3\phi_i$ 
and initial velocity $\dot\phi_i$. We use dimensionless 
parameters $\kappa_i=\lambda^3\phi_i/\rhol$ and 
$\kappa_0=\lambda^3\phi(a=1)/\rhol$ for better physical 
interpretation, and set $\Omega_{\Lambda,0}=0.7$.

All cases indeed behave as general relativity in the past.  
The no potential case has no free parameters, and shows 
strengthening gravity, $\geff>1$, with $\geff(z=0.5)=1.08$. 
It asymptotes to $\geff\to2$ in the future, where $\alb\to1$. 
The linear potential cases exhibit weakened 
gravity, with the behavior shown for a variety of initial conditions 
given by $\kappa_i$ and velocities, or field distance rolled, 
characterized by  $\kappa_0-\kappa_i$. Increasing $\kappa_i$ causes the deviation from  
general relativity to occur earlier, while increasing 
$\kappa_0-\kappa_i$ increases the deviation nearer the 
present and determines the future behavior, though they 
all eventually asymptote to $\geff\to0$ in the future (while 
$\alb\to-\infty$). However, during the observable epoch 
these models are all consistent with growth measurements, 
e.g.\ $\geff\in[0.93,1]$ for $z>0.5$. 

\begin{figure}[htb!]
\centering 
\includegraphics[width=0.6\columnwidth]{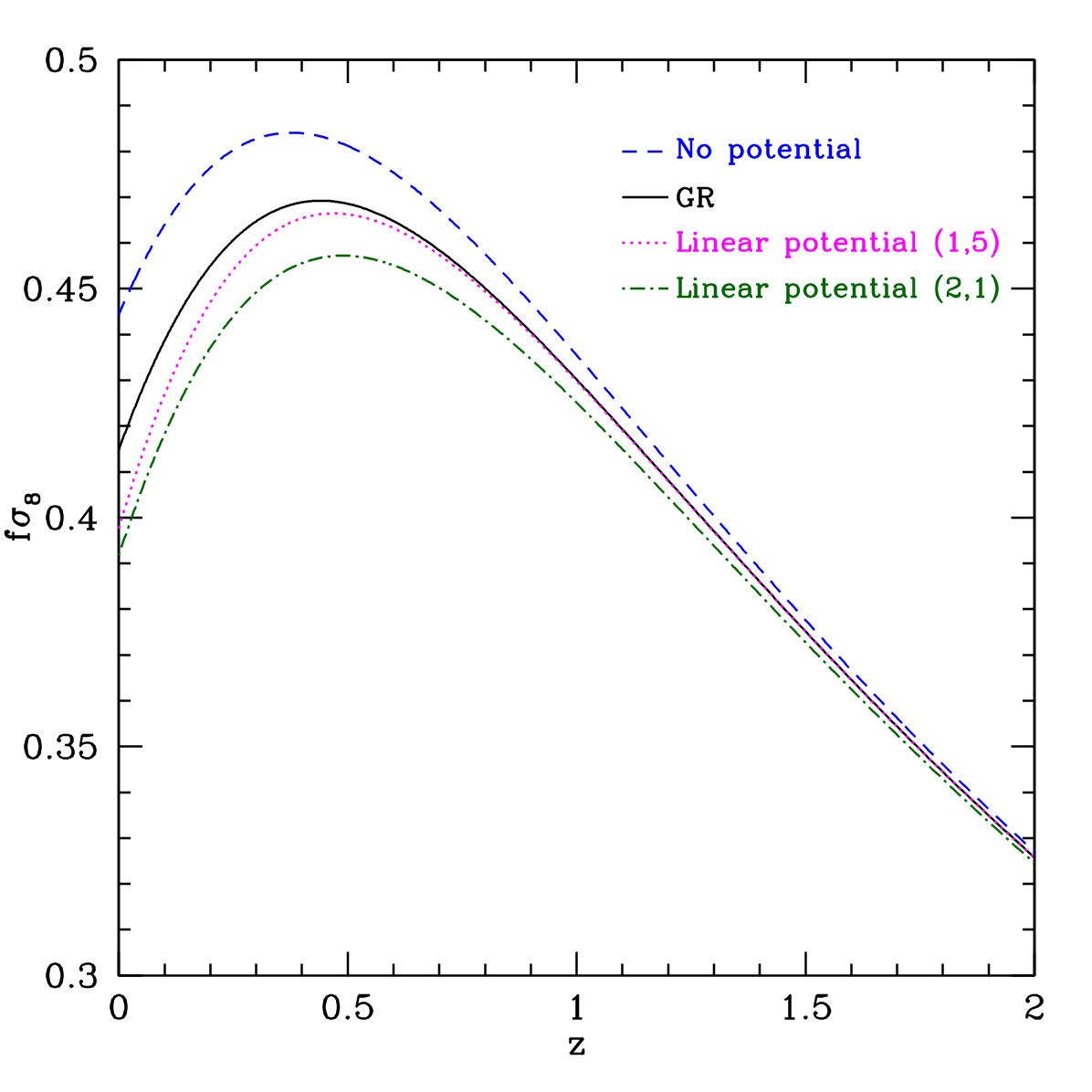}
\caption{The cosmic growth rate $f\sigma_8$ is 
plotted vs redshift, for general relativity (GR), 
the Horndessence no potential theory, and two 
Horndessence linear potential models corresponding 
to the same color curves as in Fig.~\ref{fig:geff} and 
labeled by $(\kappa_i,\kappa_0-\kappa_i)$.  
} 
\label{fig:fsig} 
\end{figure}

Figure \ref{fig:fsig} shows the growth rate $f\sigma_8(z)$ 
as can be measured from redshift space distortions in 
galaxy redshift surveys. The ongoing survey with the 
Dark Energy Spectroscopic Instrument (DESI \cite{desisci,desi19}) 
can make percent level measurements over a wide range 
of redshifts. The low redshift region where the modified 
gravity effects are strongest can gain further precision 
from peculiar velocity surveys \cite{pv1,pv2,pv3}. Note 
that the different shapes of the predicted curves help lift 
degeneracy with the amplitude of density perturbations, 
e.g.\ $\sigma_{8,0}$.

\section{Soundness} \label{sec:sound} 

The property functions $\alpha_i$ are also convenient 
for checking the soundness of the theory, specifically whether 
it is ghost free and stable to scalar perturbations. The ghost 
free condition is given by ${\cancel{g}}\equiv\alk+(3/2)\alb^2\ge0$. 

For our class of theory, 
\be  
\alk=\frac{12\dot\phi X(G_{3X}+XG_{3XX})}{H\mpl}\,. 
\ee 
For the no potential and linear potential models this becomes 
\be 
\alk=-\frac{3}{2}\olm(1+\olm)\left(1-\frac{\lambda^3\phi}{\rhol}\right)-\frac{\lambda^3\dot\phi/H}{3\mpl H^2}\,. 
\ee 
Combining this with Eq.~\eqref{eq:alb}, the ghost free condition is 
\be 
\cancel{g}=-\frac{3}{2}\olm+\frac{3}{2}\frac{\olm(1-\olm)\,\lambda^3\phi}{\rhol}-\frac{\lambda^3\dot\phi/H}{3\mpl H^2}+\frac{3}{2}\left(\frac{\olm\,\lambda^3\phi}{\rhol}\right)^2\ge0\,. \label{eq:noghost} 
\ee 

We immediately see that this is violated for the no potential ($\lambda^3=0$) 
model. For the linear potential model it can be ghost free. 
In the asymptotic future, $\phi\sim a^3$ and the last term dominates 
so the condition holds. In the asymptotic past, $\phi\sim\phi_i+a^{9/2}$, 
and 
\be 
\cancel{g}\to -\frac{3}{2}\olm\left(1-\frac{\lambda^3\phi_i}{\rhol}\right)\,. 
\ee 
So we would require the field to start from a frozen state with 
$\phi_i>\rho/\lambda^3$ to have a ghost free theory. 

To ensure stability against scalar perturbations, the sound speed 
squared must be nonnegative, 
\be 
\cancel{g}c_s^2=\frac{\alb}{2}(3\olm-1-\alb)+\alb'\ge0\,. 
\ee 
The expression evaluates to 
\be 
\cancel{g}c_s^2=\frac{5}{2}\olm\left(1-\frac{\lambda^3\phi}{\rhol}\right)-\frac{\olm^2}{2}\left(1-\frac{\lambda^3\phi}{\rhol}\right)^2
-\frac{3\olm^2}{2}\left(1-\frac{\lambda^3\phi}{\rhol}\right)-\frac{\olm}{\rhol}\frac{\lambda^3\dot\phi_i(a/a_i)^3}{H}\,. \label{eq:stable} 
\ee 
Recalling that at late times $\phi\sim a^3$, we see the second term 
dominates and $\cancel{g}c_s^2\to-(1/2)(\lambda^3\phi/\rhol)^2<0$, giving a 
late time instability. At early times $\cancel{g}c_s^2\to(5/2)\olm(1-\lambda^3\phi_i/\rhol)$ 
so we would require $\phi_i<\rho/\lambda^3$ -- the exact opposite 
of the ghost free condition! Thus neither the no potential nor the 
linear potential model is sound. 

Can we extend this no go situation to an arbitrary potential (giving up the 
desired property of shift symmetry)? For a general potential $V(\phi)$, 
the solution $G_3(X)$ will change, but we can write the $\alpha_i$ and 
the soundness conditions without solving for $G_3(X)$, just using 
Eq.~\eqref{eq:geqlinv} with $\lambda^3\phi$ replaced with $V$. 
This gives the minor change 
\bea 
\alb&=&\olm\,\left(1-\frac{V}{\rhol}\right)\\ 
\alk&=&-\frac{3}{2}\olm(1+\olm)\left(1-\frac{V}{\rhol}\right)-\frac{\dot V/H}{3\mpl H^2}\,. 
\eea  
The no ghost condition is Eq.~\eqref{eq:noghost}, simply 
with $\lambda^3\phi$ replaced with $V$ (so $\lambda^3\dot\phi\to\dot V$); 
the stability condition is Eq.~\eqref{eq:stable} with the same 
substitution (noting that $\lambda^3\dot\phi_i(a/a_i)^3=\lambda^3\dot\phi\to\dot V$). 

However, the new element is that we now have freedom 
in $\dot V$: it no longer has to go as $\dot\phi\sim a^3$. Studying 
the equations, we see that at early times  
\bea 
\cancel{g}\to -\frac{3\olm}{2}\left[1-\frac{V}{\rhol}-\olm\left(\frac{V}{\rhol}\right)^2\right]-\frac{\olm\dot V}{\rhol H}\\ 
\cancel{g}c_s^2\to \frac{5\olm}{2}\left(1-\frac{V}{\rhol}\right)-\frac{\olm^2}{2}\left(1-\frac{V}{\rhol}\right)^2-\frac{\olm\dot V}{\rhol H}\,. 
\eea 
If $V\lesssim\rhol$ then 
we need $\dot V<0$. and the magnitude must be 
$|\dot V|\gtrsim H\rhol\gtrsim HV$. But with $V$ changing rapidly 
it will eventually break the criterion $V\lesssim\rhol$. If $V\gg\rhol$ 
then if $\olm V/\rhol<1$ we need $\dot V<0$ and $|\dot V|>HV$, 
and eventually $\olm V/\rhol<1$ is overturned. If $\olm V/\rhol>1$ 
then we need $\dot V<0$ with $\olm|\dot V|/(\rhol H)>(\olm V/\rhol)^2$, or 
$|\dot V|>HV\,[V/(\mpl H^2)]$. 

At late times, 
\bea 
\cancel{g}\to \frac{3}{2}\left[\left(\frac{V}{\rhol}\right)-1\right]-\frac{\dot V}{\rhol H}\\ 
\cancel{g}c_s^2\to -\frac{1}{2}\left[\left(\frac{V}{\rhol}\right)-1\right]-\frac{\dot V}{\rhol H}\,. 
\eea 
If $V<\rhol$ then we need $\dot V<0$ and $|\dot V|>\rhol H$, 
which again will eventually become inconsistent with $V<\rhol$. 
For $V>\rhol$, we need $\dot V<0$ and $|\dot V|\gtrsim HV\,(V/\rhol)$. 
It's unclear if this can be realized: as $V$ is driven smaller by $\dot V<0$, 
it may become inconsistent with $V>\rhol$. However, if $V\to\rhol$ 
then both the ghost free and stability conditions may be satisfied. 
Another possibility is to allow a negative potential, so $V$ is 
just driven more negative. (This does not have the usual Big Crunch 
doomsday since the modified gravity term in 
Eq.~\ref{eq:geqlinv} with $\lambda^3\phi$ replaced by $V$ 
compensates, so the cosmic expansion remains as in $\Lambda$CDM.)

\section{Conclusions} \label{sec:concl} 

Horndessence generalizes scalar field evolution to achieve cosmic 
acceleration in a manner parallel to how quintessence does with a 
nonconstant potential or k-essence does with a noncanonical kinetic 
term. Here it is modified gravity through the Horndeski braiding 
function $G_3(X)$ that pushes the scalar field. While usually one 
would start with a Lagrangian function $G_3(X)$, somehow 
motivated, and derive the resulting cosmic expansion, here we 
explore the inverse path of requiring a  $\Lambda$CDM expansion 
history, as consistent with observations, and investigating the 
derived necessary function $G_3(X)$. 

The results show that even for this simple expansion behavior, 
the functional form of  $G_3(X)$ can be more involved than a 
simple a priori parametrization such as a power law or polynomial 
in $X$  -- see Eq.~\eqref{eq:g3full}! In fact, it is not even a rational 
function. Adding the simplest 
possible elaboration in terms of a linear (shift symmetric) potential 
greatly increases the complication. Thus one must take care 
when starting from parametrization of the Horndeski action 
(or effective field theory property) functions -- it is not clear that 
any measure or prior on some simple functional parametrization 
space will properly sample the observably viable cosmology. 

Horndessence as treated in this article was limited to a 
theory that preserves shift symmetry. This made it extremely 
predictive: the cosmic expansion history is that of  $\Lambda$CDM by 
construction, but the gravitational strength for matter (cosmic 
growth of structure) and light (gravitational lensing) differ 
from general relativity.  We have $\gm=\gl\ne G_{\rm  Newton}$ 
so there is no gravitational slip, and Horndessence is a type of 
No Run Gravity so gravitational wave propagation is not affected. 
General relativity holds in the early universe. We exhibit the 
braiding evolution $\alb(a)$ and $\geff(a)$, as well as the 
observable cosmic structure growth rate $f\sigma_8(a)$, for 
the two models of no potential and linear potential. For the no 
potential model there are 
the same number of free parameters as in $\Lambda$CDM. 

The two simple, shift symmetric models, with no potential and 
linear potential, however cannot satisfy the ghost free and stability 
conditions at all times. We have outlined a way around this 
by using a more general (not shift symmetric) potential. One could 
also increase the complexity of the action by allowing a $K(X)$ 
term. (Note that there is some nice exploration of this in 
\cite{nesseris}, without a potential and assuming some $X(a)$ 
or $G_3(X)$.) 
We leave such extensions to future work, but do not 
expect them to change the key result that cosmic expansion behavior 
such as $\Lambda$CDM does not lead to simple functional 
parametrizations of action functions like $G_3(X)$.

\acknowledgments 

I thank Stephen Appleby for helpful discussions. 
This work is supported in part by the Energetic Cosmos 
Laboratory and the 
U.S.\ Department of Energy, Office of Science, Office of High Energy 
Physics, under contract no.~DE-AC02-05CH11231.

\appendix

\section{$G_3(X)$ for $\Lambda$CDM Expansion} \label{sec:apxfull}

To exhibit the full complexity of a Horndeski action function such
as $G_3$ corresponding to even a simple cosmological model such as
$\Lambda$CDM expansion, we expand Eq.~\eqref{eq:g3full} fully to show
its explicit dependence on $X$: 
\bea
G_3(X)&=&G_3(X_i)-\frac{\rhol}{\rhomo}\sqrt{\frac{2\mpl}{3X_i a_i^{-6}}}\,
\left[\sqrt{\rhomo(X/X_i)^{-1/2}+\rhol}-\sqrt{\rhomo+\rhol}\right]  \notag \\
&+&\frac{\lambda^3\mpl\sqrt{2}}{\rhomo\sqrt{X_ia_i^{-6}}} 
\left\{\left[\sqrt{\rhomo(X/X_i)^{-1/2}+\rhol}-\sqrt{\rhomo+\rhol}\right]\right.\notag\\
&\qquad&\times 
\left(\phi_i+\dot\phi_i a_i^{-3}\frac{\rhomo}{2\rhol}\sqrt{\frac{\mpl}{3\rhol}}\left[\ln\frac{\sqrt{1+(\rhomo/\rhol) a_i^{-3}}+1}{\sqrt{1+(\rhomo/\rhol) a_i^{-3}-1}}-\frac{2\left[1+(\rhomo/\rhol) a_i^{-3}\right]^{-1/2}}{(\rhomo/\rhol) a_i^{-3}}\right]\,\right)\notag \\
&\qquad&-\dot\phi_i a_i^{-3}\,\frac{\rhomo}{6\rhol}\,\left[\left(\sqrt{1+(\rhomo/\rhol) a_i^{-3}(X/X_i)^{-1/2}}+1\right)\ln\left(\sqrt{1+(\rhomo/\rhol) a_i^{-3}(X/X_i)^{-1/2}}+1\right)\right.\notag\\ 
&\qquad&\qquad\left.-\left(\sqrt{1+(\rhomo/\rhol) a_i^{-3}}+1\right)\ln\left(\sqrt{1+(\rhomo/\rhol) a_i^{-3}}+1\right)\right.\notag\\
    &\qquad&\qquad-\left.\left.\left(\sqrt{1+(\rhomo/\rhol) a_i^{-3}(X/X_i)^{-1/2}}-1\right)\ln\left(\sqrt{1+(\rhomo/\rhol) a_i^{-3}(X/X_i)^{-1/2}}-1\right)\right.\right.\notag\\ 
    &\qquad&\qquad\left.\left.+\left(\sqrt{1+(\rhomo/\rhol) a_i^{-3}}-1\right)\ln\left(\sqrt{1+(\rhomo/\rhol) a_i^{-3}}-1\right)+\frac{1}{2}\ln(X/X_i)\right]\right\}\,. \label{eq:g3fulltot}
\eea


\end{document}